\documentclass[letterpaper,  
		       biblatex,
              ]{jacow}
\makeatletter%
	\ifboolexpr{bool{xetex}}
	 {\renewcommand{\Gin@extensions}{.pdf,%
	                    .png,.jpg,.bmp,.pict,.tif,.psd,.mac,.sga,.tga,.gif,%
	                    .eps,.ps,%
	                    }}{}
\makeatother

\ifboolexpr{bool{xetex} or bool{luatex}} 
 {}                                      
 {\usepackage[utf8]{inputenc}}           

\usepackage[USenglish]{babel}			 

\usepackage[final]{pdfpages}
\usepackage{multirow}
\usepackage{ragged2e}
\usepackage{xspace}

\newcommand{\htp}{\ensuremath{\mathrm{H}_2^+}\xspace}

\newcommand{\BE}[0]{\begin{equation}}
\newcommand{\EE}[0]{\end{equation}}
\newcommand{\BEA}[0]{\begin{eqnarray}}
\newcommand{\EEA}[0]{\end{eqnarray}}

\newcommand{\nuebar}{\ensuremath{\bar{\nu}_e}\xspace}
\newcommand{\nue}{\ensuremath{\nu_e}\xspace}

\newcommand{\numu}{\ensuremath{\nu_{\mu}}\xspace}

\newcommand{\nutau}{\ensuremath{\nu_{\tau}}\xspace}

\mathchardef\mhyphen="2D

\newcommand{\figref}[1]{Figure~\ref{#1}}
\newcommand{\tabref}[1]{Table~\ref{#1}}

\newcommand{\DD}{DAE$\delta$ALUS\xspace}

\begin{document}

\title{Updated physics design of the \DD and \NoCaseChange{IsoDAR} coupled cyclotrons for high intensity \htp beam production\thanks{Work supported by NSF under award \#1505858.}}

\author{D. Winklehner\thanks{winklehn@mit.edu}, 
        Massachusetts Institute of Technology, Cambridge, USA  \\
		on behalf of the \DD Collaboration.}
	
\maketitle

\begin{abstract}
The Decay-At-rest Experiment for $\delta_{\textrm{CP}}$ violation At a Laboratory for Underground Science (\DD) and the Isotope Decay-At-Rest experiment (IsoDAR) are 
proposed experiments to search for CP violation in the neutrino sector, and ``sterile'' neutrinos, respectively. In order to be decisive within 5 years, the neutrino flux and, consequently, the driver beam current (produced by chained cyclotrons) must be high. \htp was chosen as primary beam ion in order to reduce the electrical current and thus space charge. This has the added advantage of allowing for stripping extraction at the exit of the \DD Superconducting Ring Cyclotron (DSRC). The primary beam current is higher than current cyclotrons have demonstrated which has led to a substantial R{\&}D effort of our collaboration in the last years. We present the results of this 
research, including tests of prototypes and highly realistic beam simulations, which 
led to the latest physics-based design. The presented results suggest that it is feasible, albeit challenging, to accelerate \SI{5}{mA} of \htp to \SI{60}{MeV/amu} in a compact cyclotron and boost it to \SI{800}{MeV/amu} in the DSRC with clean extraction in both cases.
\end{abstract}

\section{Introduction}
\subsection{Physics Motivation}
The standard model of particle physics includes three so-called ``flavors'' of neutrinos:
\nue, \numu, and \nutau, and their respective anti-particles. These particles can change 
flavor (neutrino oscillations), a process that can be described using a mixing matrix.
This necessarily means that neutrinos must have a small mass \cite{olive:particle_review}. In addition,
some experiments aimed at measuring these oscillations in more detail have shown 
anomalies that led to the postulation of so-called ``sterile'' neutrinos which would 
take part in the oscillation, but, contrary to
the three known flavors, do not interact through the weak force \cite{collin:sterile}. 
Another important question is whether the three neutrino model can give rise to a CP-violating phase
$\delta_{\textrm{CP}}$ \cite{mena:cp_violation}, 
which might explain the matter-antimatter 
asymmetry in the universe today.
\DD \cite{abs:daedalus, aberle:daedalus} and IsoDAR \cite{bungau:isodar} are 
proposed experiments to search for CP violation in the neutrino sector, and 
sterile neutrinos, respectively.
In the following, we will give a brief overview of the facilities and identify 
and discuss the most critical aspects.

\newpage
\subsection{Facilities Overview}
\begin{figure}[t!]
	\centering
		\includegraphics[width=0.8\columnwidth]{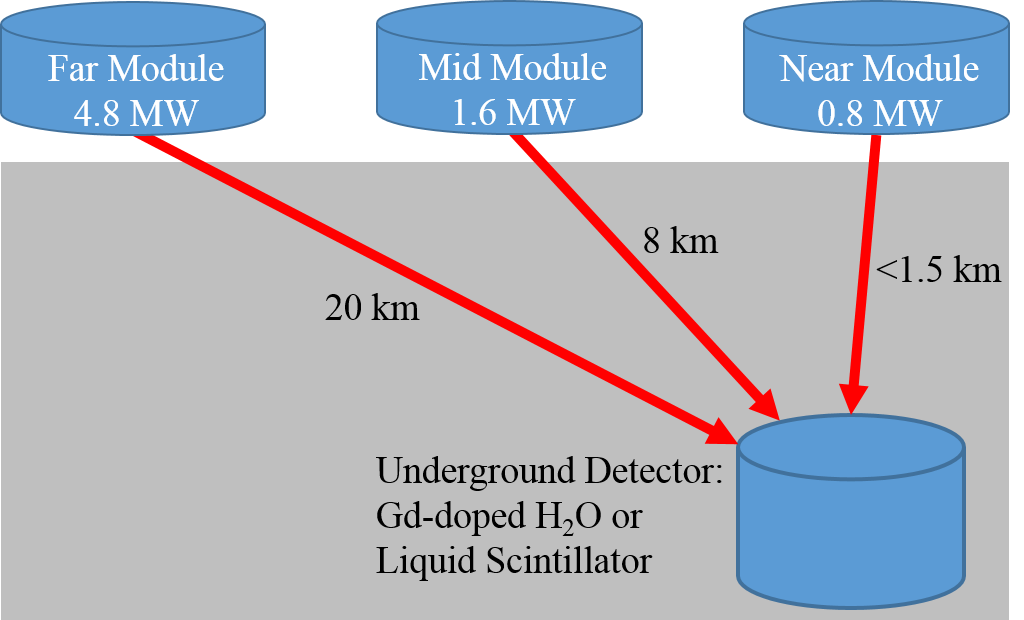}
		\caption{Schematic of the layout of \DD accelerator modules. 
		         The powers at the respective modules, are average
	             values based on a 20\% duty cycle.
	             \label{fig:dd_concept}}
\end{figure}
\begin{figure}[t!]
	\centering
		\includegraphics[height=0.19\textheight]{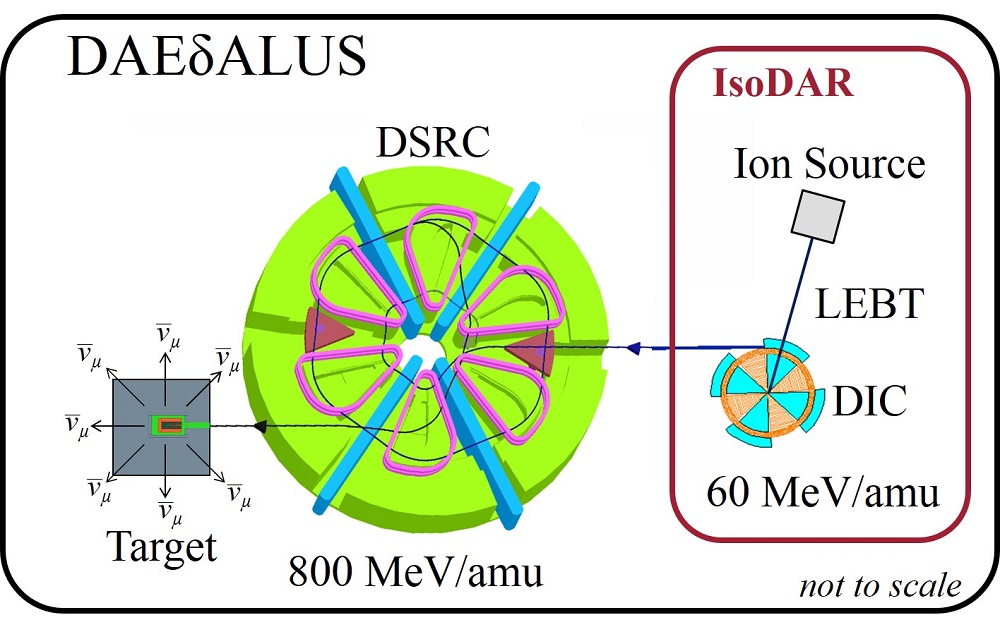}
		\caption{Cartoon picture of one single \DD module with the 
				 injector part that can be used for IsoDAR highlighted
				 on the right.
	             \label{fig:dd_isodar}}
\end{figure}

In the \DD concept (described in detail in \cite{abs:daedalus, aberle:daedalus}), 
three accelerator modules are placed at distances 1.5, 8, and \SI{20}{km} from 
a large detector (see \figref{fig:dd_concept}). 
As the neutrino oscillation probability depends on $L/E$, the ratio of neutrino 
energy to the distance traveled \cite{olive:particle_review},
this scheme can work as follows:
The near module constrains the flux, the mid module constrains the rise of 
the probability wave and the far module measures the oscillation maximum. 
Each of these modules 
consist of one or more chains of cyclotrons, as depicted in \figref{fig:dd_isodar}.
The neutrino distribution from the production target is more or less isotropic, 
which means the number of produced neutrinos needs to increase with distance
if one wants to keep statistics up. Hence
the higher power of the far site which will be reached by using several modules.
In this way, \DD can be used to measure a $\delta_{\textrm{CP}}$ dependent 
maximum of the oscillation curve.
\figref{fig:dd_isodar} shows schematically the main parts of \DD:
\begin{enumerate}
\vspace{-0.5\topsep}
\itemsep-0.2em
\item Ion source
\item Low Energy Beam Transport (LEBT)
\item \DD Injector Cyclotron (DIC)
\item Medium Energy Beam Transport (MEBT)
\item \DD Superconducting Ring Cyclotron (DSRC)
\item High Energy Beam Transport (HEBT)
\item Neutrino production target
\end{enumerate}
As \DD is a big project, it makes sense to look for a staged approach and
physics that can be done with only part of it. 
In this case using only the DIC and replacing the DSRC with a different 
production target comes to mind. This is IsoDAR, a search for sterile
neutrinos. Here the primary \htp beam at \SI{60}{MeV/amu} is used
to produce \nuebar through isotope-decay-at-rest.
In both experiments, the primary ion beam current needs to be very high in order 
for the measurements to be conclusive within a certain time span. 
The nominal current is \SI{5}{mA}. IsoDAR will operate with a 90\% 
duty cycle, \DD with 20\%. Clearly the target design is a major challenge for
beam powers of ~\SI{600}{kW} and ~\SI{1.6}{MW}, respectively. However, for
the sake of brevity and to keep with the topic of the conference,
we will abstain from a discussion of the targets and
instead point to the references given throughout this text. 
Similarly, we will not consider the MEBT and HEBT here as they are not 
considered high-risk. Instead, in the following section, we will discuss 
the ion source, LEBT, DIC and DSRC in more detail (henceforth called the 
``driver'').

\section{Motivation of Design Parameters}
As the isodar front-end and cyclotron are, for all intents and purposes,
identical to the \DD front-end and DIC, we will discuss IsoDAR first and
consider only the DSRC in the subsequent \DD section.

\subsection{IsoDAR}
\begin{table}[b!]
    \caption{Important parameters for the IsoDAR driver.\label{tab:isodar}}
    \centering
    \renewcommand{\arraystretch}{1.1}
    \begin{tabular}{llll}
        \textbf{Parameter} & \textbf{Value} & \textbf{Parameter} & \textbf{Value}\\
        \hline
        Ion & \htp & Ion Source & Multicusp \\
        $\mathrm{I}_{\textrm{source, nom.}}$ & \SI{35}{mA} DC & Injection & Spiral Infl.\\
        Cycl. Freq. & \SI{42.1}{MHz} & Harmonic & 6  \\
        Cycl. Type & Compact & $\mathrm{E}_{\textrm{max}}$ & \SI{60}{MeV/amu} \\
        Extraction & Septum & $\mathrm{I}_{\textrm{cycl., extr.}}$ & \SI{5}{mA} avg.\\
        \hline
    \end{tabular}
\end{table}
Recently, the \DD Collaboration published a Conceptual Design Report (CDR) for
the technical aspects of the IsoDAR project \cite{abs:isodar}, in which 
the project is discussed in much detail. The important parameters of the 
front end and cyclotron are summarized in \tabref{tab:isodar}. From beginning to
end, the driver consists of 1.) an ion source, 2.) a LEBT with buncher, and 3.)
a compact isochronous cyclotron. \htp was chosen as primary beam ion because
of the reduction in electrical beam current vs. particle current (after stripping), 
which reduces space charge effects. At the same time the magnetic rigidity of the
beam increases, which has to be taken into account during injection.
The main challenges were identified as
\begin{enumerate}
\vspace{-0.5\topsep}
\itemsep-0.2em
\item Production of necessary initial current in the ion source.
\item Beam injection into the cyclotron through a spiral inflector
of appropriate size to accommodate high rigidity.
\item Focusing and matching in the first 10 turns.
\item Ultra-low-loss extraction from the cyclotron.
\end{enumerate}
\subsubsection{About items 1. and 2.}
In the summers of 2013 and 2014 we tested \htp ion production, 
LEBT and cyclotron injection in collaboration with Best 
Cyclotron Systems, Inc. (BCS) . The results were published in
\cite{winklehner:bcs_tests} and can be summarized as such:
An off-resonance ion source like the one we tested (Versatile Ion Source - VIS) can provide the necessary \htp ion flux, but only marginally and through pushing the source to its limits.
Consequently, we are now investigating alternatives to the VIS 
and conventional LEBT system. We are pursuing two
avenues: 
\begin{itemize}
\vspace{-0.5\topsep}
\itemsep-0.2em
\item We are currently building a new multicusp ion source
at MIT called MIST-1 \cite{axani:mist1}, which
is optimized for \htp.
\item We are investigating the use of an RFQ to directly inject a highly bunched beam into 
the spiral inflector \cite{winklehner:rfq1}.
Funding for a first RFQ injector was obtained and the first 
phase (design study) will commence this fall.
\end{itemize}
Furthermore, during the BCS tests, we could show that a large
(\SI{1.6}{cm} gap) spiral inflector could be built and
operated at up to $\pm$\SI{12}{kV}.
\SI{6}{mA} of a DC \htp beam were injected through the spiral
inflector and results compared well with  simulations.

\subsubsection{About items 3. and 4.}
\begin{figure}[t!]
	\centering
		\includegraphics[height=0.19\textheight]
		                {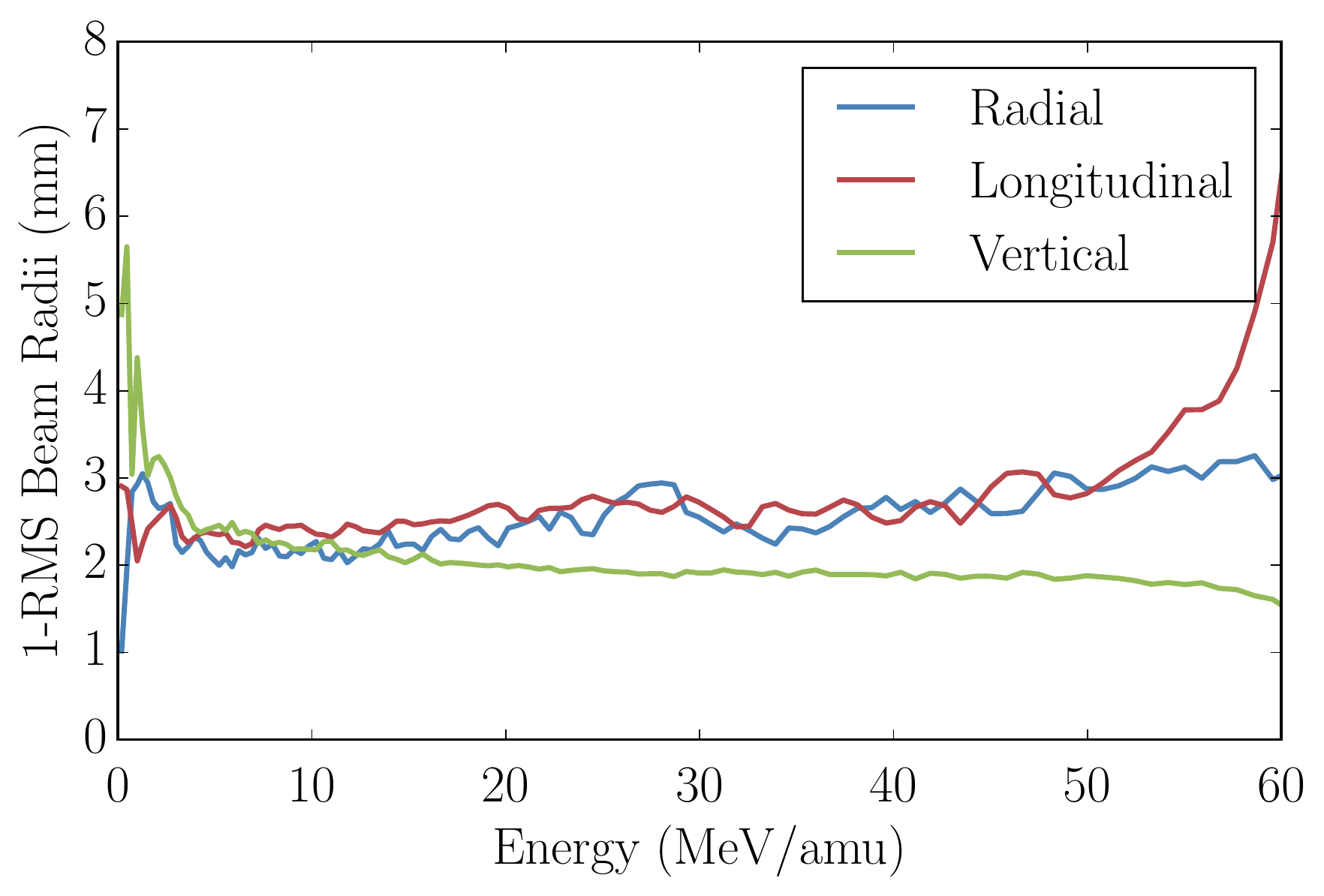}
		\caption{1-RMS beam radii for a matched beam in IsoDAR,
		         starting at \SI{193}{keV/amu}. 
		         Collimators are placed
		         in turn 10 ($\approx$\SI{3}{MeV/amu}) which  
		         clean up beam halo, and only reduce the beam 
		         current from \SI{6.5}{mA} to \SI{5.0}{mA}.
		         Data from \cite{jonnerby:thesis}. 
	             \label{fig:isodar_rms}}
\end{figure}
\begin{figure}[t!]
	\centering
		\includegraphics[height=0.16\textheight]
		                {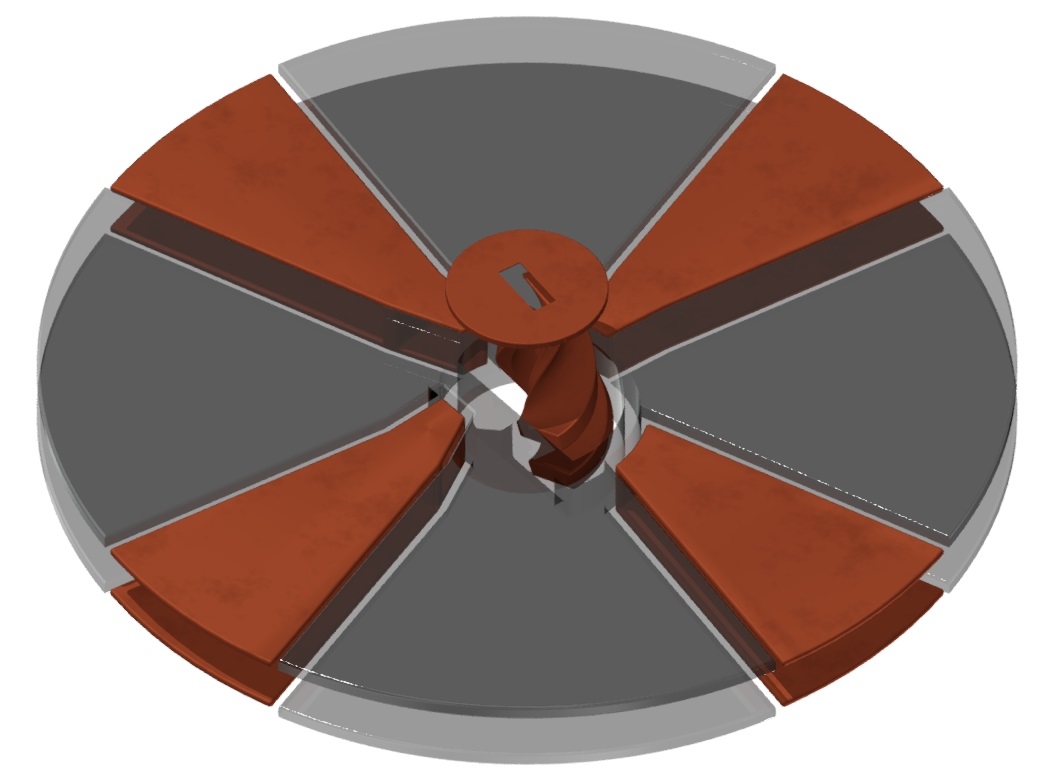}
		\caption{Preliminary CAD model of the IsoDAR central
		         region.
	             \label{fig:isodar_cr}}
\end{figure}
In a previous publication, it was shown through particle-in-cell
(PIC) simulations using OPAL \cite{adelmann:opal} that, starting
at \SI{1.5}{MeV/amu}, a stationary distribution in the horizontal plane could be achieved through vortex-motion
\cite{yang:daedalus}.
Through collimation at low energy and tuning of the RF phase,
it was possible to keep the predicted beam loss on the 
septum below \SI{200}{W}.
In the past few months, these simulations were extended to
lower energies and matched distributions were found down
to energies of \SI{193}{keV/amu}. An example is shown in 
\figref{fig:isodar_rms}. Note that the steep increase in
longitudinal
direction at the end stems from a resonance, which is expected
to be suppressed in future design iterations of the magnetic 
field. The vertical beam size is fairly large in the first 
few four turns and then decreases rapidly. We are currently in
process of designing a central region that can accommodate this
large beam. The present state of the design is depicted in
\figref{fig:isodar_cr}.

\subsection{\DD}
\begin{table}[b]
    \caption{Important parameters for the \DD driver. 
             Note that the DIC parameters are identical
             to the IsoDAR parameters listed in 
             \tabref{tab:isodar}.
             \label{tab:daedalus}}
    \centering
    \renewcommand{\arraystretch}{1.1}
    \begin{tabular}{llll}
        \textbf{Parameter} & \textbf{Value} & \textbf{Parameter} & \textbf{Value}\\
        \hline
        Ion & \htp & Injection & Radial\\
        Cycl. Freq. & \SI{42.1}{MHz} & Harmonic & 6  \\
        Cycl. Type & Ring & $\mathrm{E}_{\textrm{max}}$ & \SI{800}{MeV/amu} \\
        Extraction & Stripping & $\mathrm{I}_{\textrm{cycl., extr.}}$ & \SI{5}{mA} avg.\\
        \hline
    \end{tabular}
\end{table}

The \DD design was reviewed in detail in \cite{abs:daedalus}
and the most important parameters of the DSRC are listed in 
\tabref{tab:daedalus}. The superconducting ring cyclotron 
will take the \SI{60}{MeV/amu} beam from the IsoDAR-like
front-end and boost it to \SI{800}{MeV/amu}. Detailed simulations
were performed and the results published in \cite{yang:daedalus}.
These simulations showed that a stationary distribution 
forms in the DSRC which can then be extracted very cleanly 
through stripping extraction.
\begin{figure}[t!]
	\centering
		\includegraphics[height=0.16\textheight]
		                {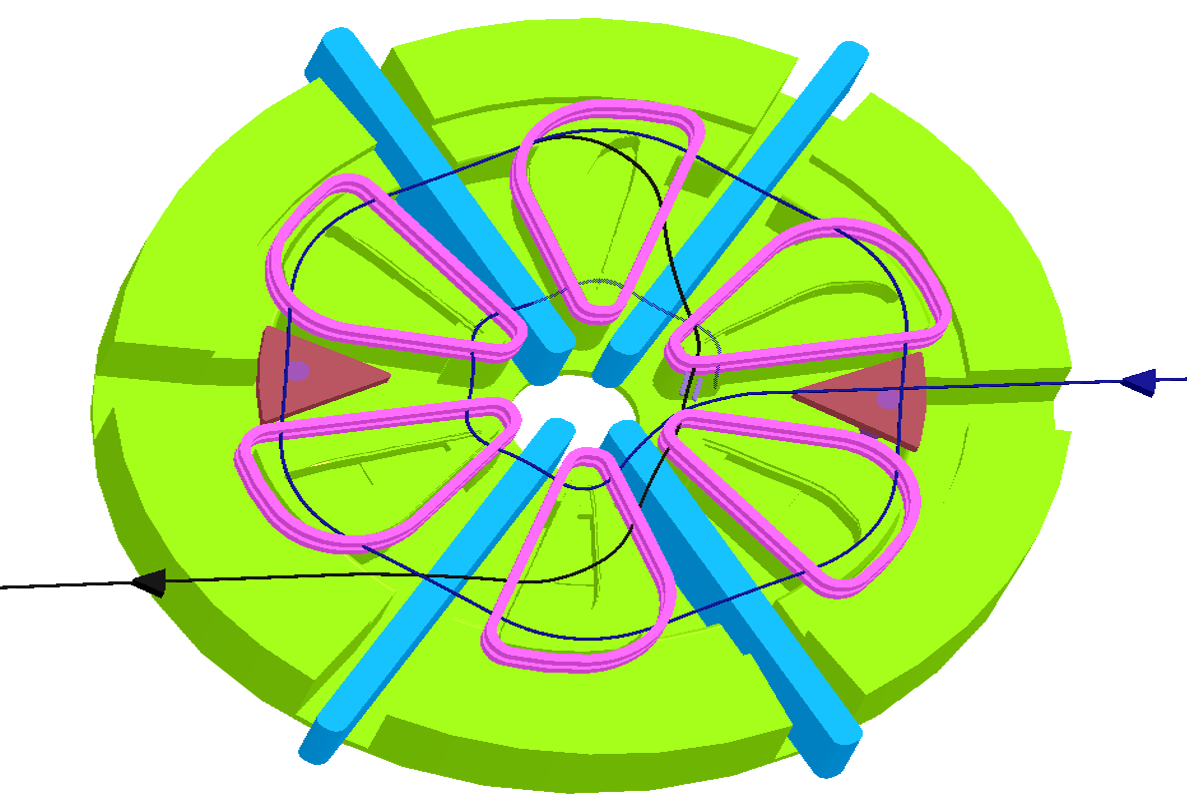}
		\caption{Preliminary CAD model of the DSRC with injection
		         and extraction trajectories.
	             \label{fig:dsrc}}
\end{figure}
A model of the current design of the DSRC can be seen in 
\figref{fig:dsrc}.

\section{Conclusion}
\DD and IsoDAR are ambitious experiments aiming at discovering CP violation
in the neutrino sector and the existence of sterile neutrinos, respectively.
The requirement of \SI{5}{mA} of \htp has led to a substantial R\&D effort 
on which was reported here. We have identified the injection of the necessary 
current into the compact DIC (almost identical to the IsoDAR main cyclotron)
as the main challenge and have shown preliminary experimental studies and 
simulations that will soon lead into a full start-to-end simulation treatment
of the system. The experimental high intensity injection studies showed 
that a spiral inflector with the required large size for the higher magnetic
rigidity of the \htp beam can be built and operated. 
Simulations using OPAL with the new spiral 
inflector option compared well to these studies and systematic injection 
simulations using beam currents up to the required injection currents are 
on the way. Parallel to the conventional LEBT front end, we have just begun
a full investigation of using an RFQ for direct axial injection of a bunched
beam into the spiral inflector.

\printbibliography
\newpage

\end{document}